\documentclass[%
 reprint,
 amsmath,amssymb,
 aps,
]{revtex4-1}

\usepackage{dcolumn}
\usepackage{bm}
\usepackage{graphicx}
\usepackage{caption}
\usepackage{refstyle}
\usepackage{amsmath}
\usepackage{amssymb}
\usepackage[mathlines]{lineno}

\begin{document}


\title{Monolayer Adsorption of  Noble Gases on Graphene}

\author{Sidi M. Maiga}
\author{Silvina M. Gatica}%
 \email{sgatica@howard.edu}
\affiliation{  Department of Physics and  Astronomy, Howard University \\
 2355 Sixth St NW, Washington, DC 20059, USA }

\date{\today}

\begin{abstract}
We report our results on the adsorption of noble gases such as argon, krypton and xenon on a   graphene sheet, using Grand Canonical Monte Carlo (GCMC) simulations.    
We calculated the two-dimensional gas-liquid critical temperature for each adsorbate, resulting in fair agreement with    theoretical  predictions and experimental values of gases on graphite. 
We    determined the different phases of the monolayers  and constructed the phase diagrams. We found    two-dimensional incommensurate solid phases for krypton, argon and xenon, and a two-dimensional commensurate solid phase for  krypton.

\end{abstract}

\pacs{Valid PACS appear here}
\maketitle


\section{\label{intro}Introduction}

The properties of adsorbed monolayers (one-atom thick layer) have been investigated for many years, motivated by the realization of 2D matter and the implications on the new technologies based on new substrates. For example, adsorption of noble gases have been studied on many substrates like graphite \cite{damico1990,larher1974,migone1984,LarherGilquin1979,thomy1981} and    metals \cite{unguris1981,kern1991,ramseyer1994,glachant1981}. 

Bruch et.al. constructed the phase diagrams of noble gases adsorbed on  graphite, by collecting data from several experiments. \cite{bookch6}  For monolayer Kr,  two solid phases   are observed in the diagram: an incommensurate solid (IS) and a commensurate solid (CS) with a fractional density of 1/6  (one Kr atom per every 6 carbons).  Monolayer Ar and Xe, on the other hand, do not exhibit a 2D- CS phase. 

More recently in Ref. \cite{bruch2010} , Bruch. et.al. studied the  case of films formed on both sides of a suspended graphene sheet.  Superfluid helium  on graphene got particular attention  as well. \cite{boninsegni2013,gordillo2009,ceperley2012}

We have studied the behavior of Kr  and Ar on   suspended single-walled carbon nanotubes 
\cite{kim2011,Mbaye2016}, and found that Kr forms a  cylindrical shell that has a unique CS  structure with fractional density 1/4. The 1/4-CS appears exclusively on medium-sized zigzag nanotubes.
On larger NTs, Kr forms a shell of  coverage 1/6 that is, however, not solid.
  For Ar on the other hand we have not observed commensuration in  any  of the nanotubes tested.

In this paper we report numerical results of the monolayer adsorption of   Ar, Kr and Xe   on a graphene sheet, placed at the bottom of the simulation cell. We use the method of  Grand Canonical Monte Carlo (GCMC) simulation. GCMC allows us to calculate the average number of atoms adsorbed on graphene as a function of the pressure of the vapor and temperature of the system.  In the simulations we  collect sample configurations that we use to test the structure of the monolayers.  From  calculations of the radial distribution function and structure factor we evaluate the phase of the monolayer (liquid, IS or CS).

This paper is organized as follows. In the next section, we present the methodology, the model for the interaction potentials, and the GCMC simulation method. In section \ref{results} we describe the results  of the adsorption of Ar, Kr, and Xe and discuss the structure of the monolayers. Finally in section \ref{summary} we summarize and present our conclusions.


\section{\label{method} Methodology}

We used the method of GCMC, \cite{frenkel2001understanding}  in which  the average  number of particles and energy of the system is computed. The method has been extensively used for simulations of adsorption. \cite{calbi2008,calbi2001,gatica2001quasi} 

The simulation cell is a rectangular box of  size  $X = 39.35$ \AA, $Y = 38.46$ \AA, and $Z = 70$ \AA. 
The graphene sheet was positioned at the base of the simulation cell ($Z = 0$) so that the adsorption occurs only on one side of it.

The boundary conditions were set periodic in the $X$ and $Y$ directions and reflective in $Z$.  

For each gas,  simulations were run at several values of the temperature  and pressure of the vapor. For Ar, the simulation temperature ranged from 41 K to 79 K at 2 K increments;  from 70 K to 140 K with an increment of 5 K for Kr and from 90 K to 160 K at 5 K increments for Xe. Each GCMC simulation was done at fixed temperature ($T$) and chemical potential ($\mu$); $\mu$ is related to the pressure of the vapor in equilibrium with the adsorbate by the equation of state. In our simulations we considered the vapor an ideal gas.

  The graphene layer was assumed to be rigid and infinitely wide on the $X,Y$ plane, which we realized by setting periodic boundary conditions on the $X,Y$  directions. The input data of the simulation are the pressure of the vapor ($P$), and the temperature ($T$); the output data are the average number of adsorbed atoms ($N$), the averages of the total energy ($E_T$), the energy gas-surface ($E_{gs}$) and the energy gas-gas ($E_{gg}$). We also collected  samples of the coordinates of the adsorbed atoms. The number of  MC moves for each single data point in the isotherms $N(P,T)$ was  typically  $3\times 10^6$, to reach equilibrium. Additionally,  $1\times 10^6$  moves  were  performed for data collection. For each  temperature, the simulation was run 90  times;  each time, the pressure was increased by 10 \%. The ratio of creation/deletion/translation steps was set to  0.40/0.40/0.20.

\begin{figure}[p]
  \includegraphics[width=\columnwidth]{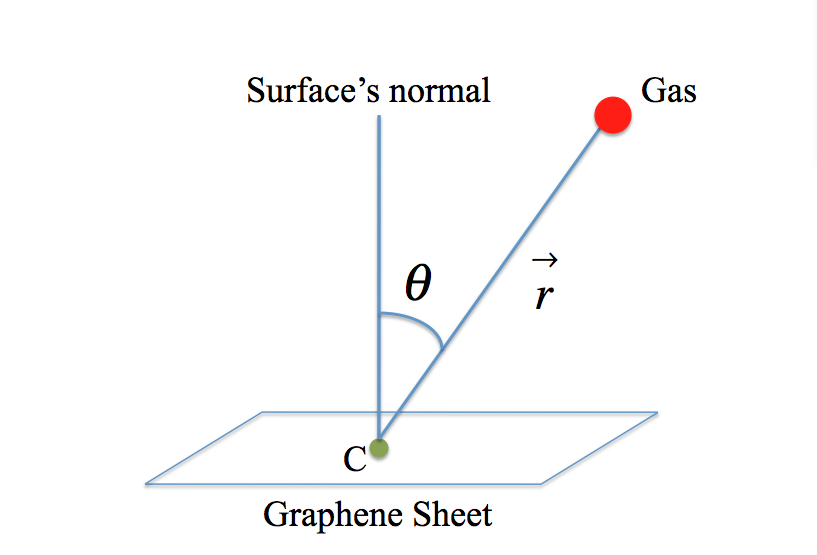}
  \caption{Schematic view of a gas molecule on top of the graphene sheet.} 
  \label{fig:mol}
   \end{figure}

The potential interaction energy between the adsorbate and the graphene was calculated with the Lennard Jones (LJ) anisotropic potential.\cite{carlos1980interaction} In the anisotropic potential, the pairwise interaction depends not only on the distance between the carbon atoms and the gas molecules, but also on the angle between the vector $\vec{r}$ (relative position of   the gas atom and the carbon atom) and the surface's normal (see \Figref{mol}). The potential is  given by,

\begin{align}
U(\vec{r}) = & 4\epsilon  
 \Big \{ \Big(  \frac{\sigma}{r}   \Big )  ^{12}  
\Big[ 1  + \gamma_R \big( 1 - \frac{6}{5} \cos^2{\theta} \big) \Big]- \\
  &\Big( \frac{\sigma}{r}\Big)^6 
\Big[ 1 
+ \gamma_A \big( 1 - \frac{3}{2} \cos^2{\theta}\big)\Big] \Big \} , \nonumber
\end{align}
where $\epsilon$ and $\sigma$ are the LJ parameters and $r$ is the distance between the gas atom and the carbon atom. $\gamma_A$ and $\gamma_R$ determine the anisotropy of the dispersion potential  and $\theta$ is the angle between $\vec{r}$ and the surface\textsc{\char13}s normal as shown in \Figref{mol}. This angular dependence originates in the anisotropy of the $\pi$ bonds of the C atoms in the graphene sheet; as a result the potential becomes more corrugated.

  The LJ parameters for the adatom-C potential are obtained by fitting physical properties of the gases\cite{stan2000} and using the semi-empirical Lorentz-Berthelot combining rules: \cite{scoles1990}
  
 \begin{align}
 \sigma_{aC} &= \frac{\sigma_{aa} + \sigma_{CC}}{2} \\
 \epsilon_{aC} &= \sqrt{\epsilon_{aa} \epsilon_{CC}}
 \end{align}

\begin{table}[!h]
\begin{tabular}{|c|c|c|} \hline
Adsorbate & $\epsilon_{aa}$(K) & $\sigma_{aa}$(\AA) \\ \hline
Ar& 120.0 & 3.4 \\ \hline
Kr & 171.0 & 3.6 \\ \hline
Xe & 221.0 & 4.1 \\ \hline
C (graphene) & 28.0 & 3.4 \\ \hline
\end{tabular}
\caption{Lennard Jones parameters of the adsorbates and substrate  \cite{stan2000}}
\label{tab:LJpara}
\end{table}

For our study, we have adopted $\gamma_A = - 0.54$  and $\gamma_R = 0.38$   based on a previous study of He on graphite by Cole et al. \cite{carlos1980interaction}.

The interaction energy among adsorbate's atoms was calculated with the regular LJ potential.

To determine the phase behavior of the adsorbed monolayers, we inspected the discontinuities or steps in the energy gas-surface function $E_{gs}(P)$ and the adsorption isotherms $N(P)$. We carefully analyzed sample configurations around  steps in the isotherms. Based on the radial distribution function, we determined the phase (solid or liquid). Similarly, we evaluated the commensuration i.e.  CS or IS based on the structure factor. 

The two-dimensional  radial distribution function (2D-RDF) $g(r)$ of a system of particles  describes how the density varies as a function of the  distance from a reference particle. For a distribution of atoms or molecules on a surface, it is given by the following equation:

\begin{align}
g(r) = \frac{N_r}{2\pi dr{\rho}^2},
\end{align}
where $N_r$ is the number of pairs of atoms at a distance between $r$ and $r+dr$ and $\rho$ is the number of atoms per unit area.

The structure factor $S(\vec{k})$ characterizes the amplitude and phase of a wave diffracted from crystal lattice planes. It is simply the squared modulus of the Fourier transform of the pair correlation function given by,

\begin{align}
S(\vec{k}) = {\left| \frac{1}{N} \sum\limits_j^N \exp\left( i\vec{k} \cdot \vec{R_j}\right) \right|}^2,
\end{align}
where $\vec{k}$ is the 2D wave vector and $\vec{R_j}$  is the  position vector of the $j^{th}$ particle in the monolayer. $\vec{k}$ can be written in terms of the reciprocal lattice vectors $\vec{b_1}$ and $\vec{b_2}$, 

\begin{align}
\vec{k} = m_1\vec{b_1} + m_2\vec{b_2},
\end{align}
with $m_1$ and $m_2$ arbitrary real numbers. For a perfect lattice, $S(\vec{k})$ is equal to 1 for all integer $m_1$ and $m_2$, and  identically zero for non-integer values. On the other hand, $S(\vec{k}\neq 0) \ll 1$  for an array of atoms in a non-matching lattice or a non-solid phase.



\section{\label{results}Results}
\subsection{\label{sec:level3}Argon}

\begin{figure}[p]
 \includegraphics[width=5in]{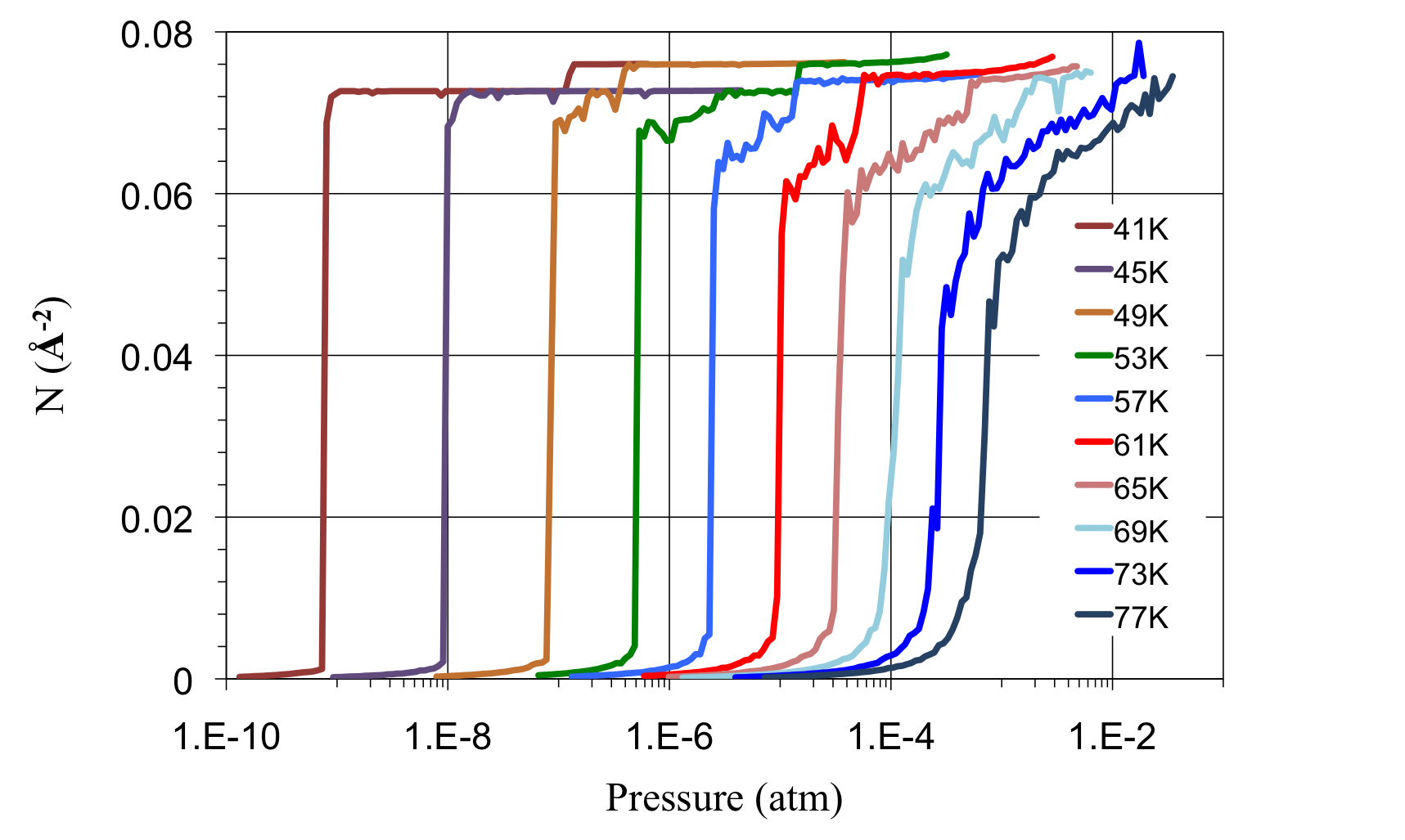}
  \caption{Adsorption Isotherms of Ar on graphene at temperatures from 41 K to 77 K, from left to right} 
  \label{fig:Ar}
   \end{figure}

\Figref{Ar} represents the adsorption isotherms $N(P)$ for argon adsorbed on   graphene.   For each temperature, we observe that at low vapor pressure the amount of  adsorbed atoms     is very low ($N\sim 0.001$ \AA$^{-2}$), which corresponds to a 2D gas.  At a higher pressure, the density of the film rapidly raise. The lower the temperature the faster the film's density grows. For example, at the lowest temperature seen in \Figref{Ar} (41 K), the isotherm steps straight up   to form a film of density 0.0726 {\AA}$^{-2}$.   Vertical  steps in the isotherms appear for higher temperatures as well, up to the critical temperature $T_c$ at which the isotherms qualitatively change to  smooth continuously growing  curves. For temperatures lower than $T_c$, there is a coexistence of a 2D gas and a dense monolayer, which can be solid or liquid. As the temperature increases, the  gap between the gas and monolayer densities  decreases approaching zero at $T=T_c$. The critical temperature $T_c$ is computed as the temperature at which  the inverse slope of the isotherms   becomes zero.   The result of our calculation is $T_c = 61$ K $\pm 1$ K.

The critical temperature of a two-dimensional Lennard-Jones lattice has been   theoretically predicted to be   $T_{c(LJ)} =0.52 \epsilon$. \cite{LJTC} In the case of Ar it results    62.4  K, in excellent agreement with our calculation. The critical density $n_{c(LJ)}$ is also predicted theoretically to be $0.35 \sigma^{-2}$, which for Ar is 0.03 \AA$^{-2}$, consistent with our simulations.  Experimental measurements of the critical point of monolayer Ar on graphite are $T_c=59$ K \cite{LarherGilquin1979} and 55 K \cite{migone1984}, which are 11\%  lower than the theoretical or simulation predictions.

By inspecting the 2D-RDF and the structure factor of several samples in the monolayer regime we  determined the phases  (G, L, IS or CS)

At the lowest temperatures, the density remains constant for a wide range of pressures, seen as the flat portion of the isotherm. At higher  vapor pressure, the monolayer slightly compresses from 0.0726 to 0.076 {\AA}$^{-2}$. In this regime, the argon monolayer is an incommensurate solid. At higher temperature, on the other hand, the monolayer is a 2D liquid transitioning to a 2D IS. Hence, for Ar we found three phase transitions: 2D-G  to 2D-L, 2D-G to 2D-IS and 2D-L to 2D-IS.

For comparison,  the 2D-density of the commensurate solid phase on graphene or graphite, which has a ratio  1/6 adsorbate/carbon atoms is   0.063 \AA$^{-2}$ while the 2D-LJ solid lattice has a density  0.079 \AA$^{-2}$.  In our simulation, we obtained a monolayer of density 0.0726  - 0.076 \AA$^{-2}$, consistent with an incommensurate solid phase. 

\begin{figure}[p]
  \includegraphics[width=5in]{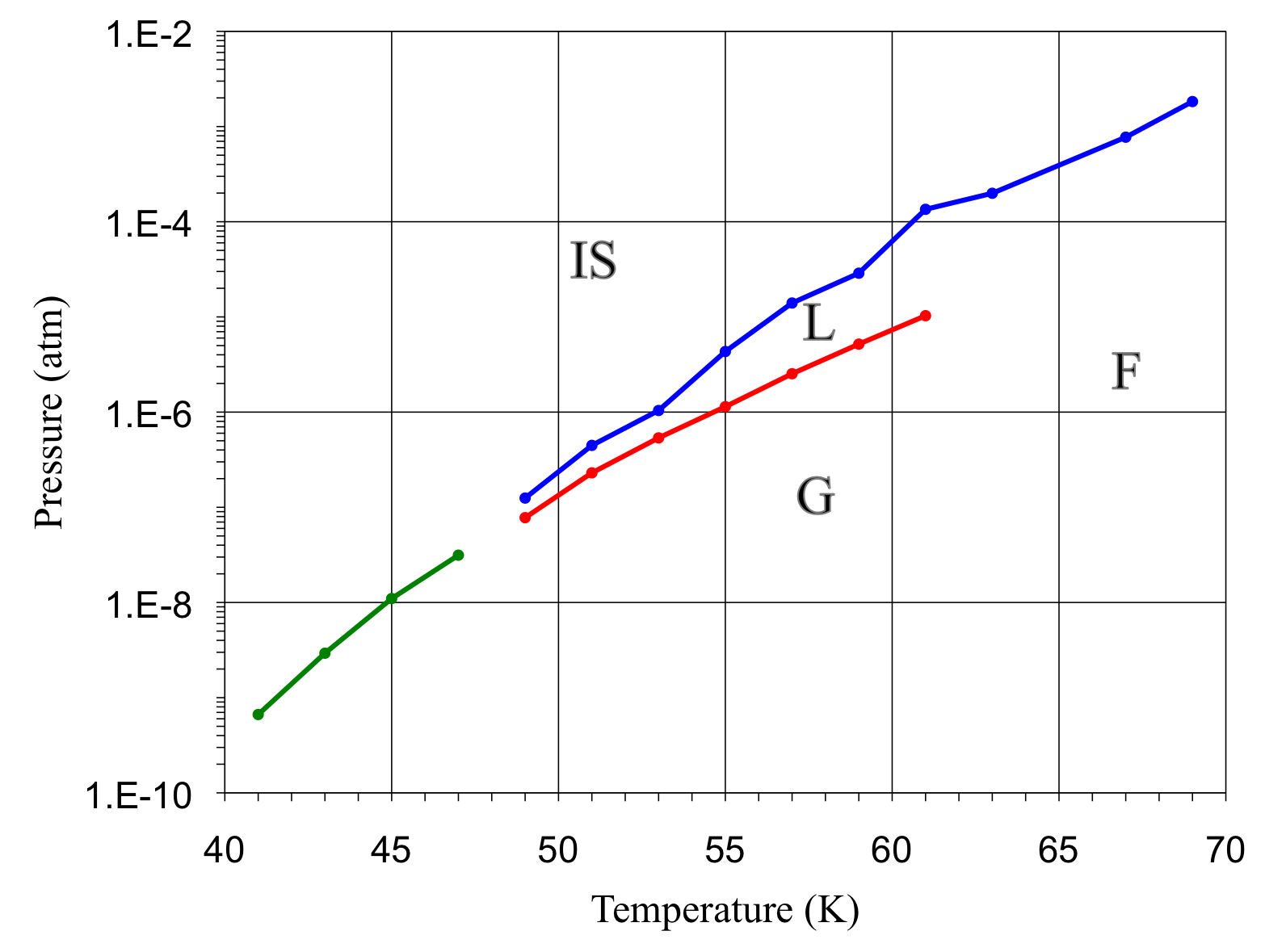}
  \caption{Phase diagram of monolayer Ar  on graphene. Labels indicate  incommensurate solid (IS), liquid (L), gas (G) and fluid (F)  phases. } 
  \label{fig:phase-Ar}
   \end{figure}

Our results are summarized in a phase diagram, shown in \Figref{phase-Ar} where we see  three coexistence lines:  G-IS and G-L  and  L-IS. We   estimated the triple point temperature at  $T_t = 47$ K $\pm 1$ K. This value is in excellent agreement  with  the experimental report for Ar on graphite from Migone et al. \cite{migone1984} ($T_t=47$ K) and in fair agreement with  D'Amico et al. \cite{damico1990} ($T_t=49.7$ K).

\subsection{\label{sec:level3} Krypton }

\begin{figure}[p]
  \includegraphics[width=5in]{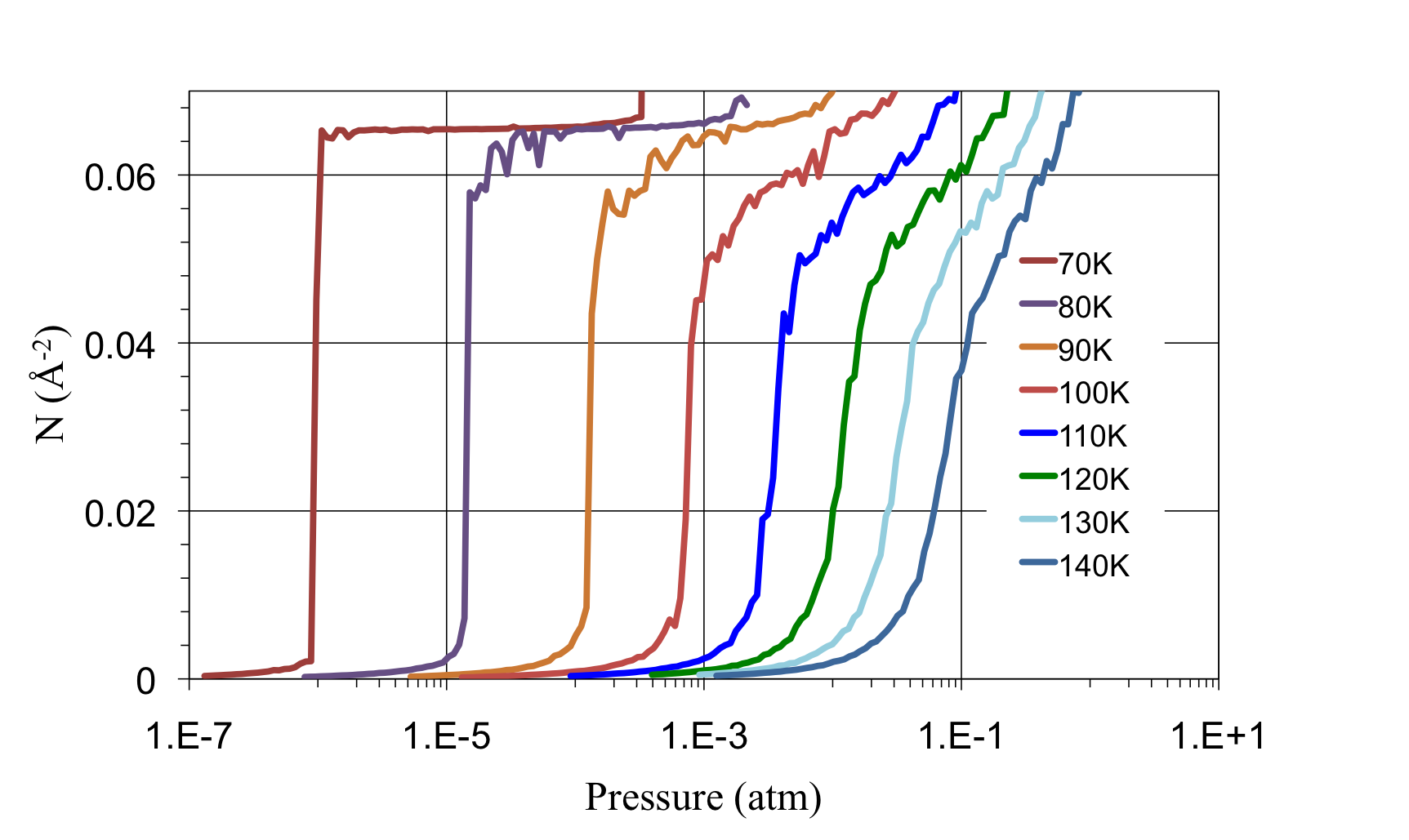}
  \caption{Adsorption Isotherms of Kr on graphene at temperatures 70K, 80K, 90K, 100K, 110K, 120K and 130K from left to right} 
  \label{fig:Kr}
   \end{figure}

We  studied the monolayer adsorption of Kr on graphene with the same methodology. The adsorption isotherms for few temperatures are shown in \Figref{Kr}.
 Using the same analysis as in the argon   case, we estimated a 2D G-L critical   temperature $T_c = 90$ K $\pm 5$ K, which agrees with the theoretical value $T_{c(LJ)} = 89 $ K and is 6\% higher than the experimental result reported in ref. \cite{LarherGilquin1979} (85.3 K).
The theoretical prediction of the 2D-LJ critical density is  $n_{c(LJ)} = 0.027$ \AA$^{-2}$, consistent with our simulations. 

 The monolayer density of Kr in the simulation is 0.065 \AA$^{-2}$,  which is lower than the density of the  2D-LJ lattice (0.071 \AA$^{-2}$) and similar to the 1/6 CS (0.063 \AA$^{-2}$).  Hence, contrary to the case of Ar,  Kr forms both commensurate and incommensurate solid phases.  The CS phase is observed at low temperature and low pressure, while the IS appears at higher $P$ and $T$.

\begin{figure}[p]
\includegraphics[width=6in]{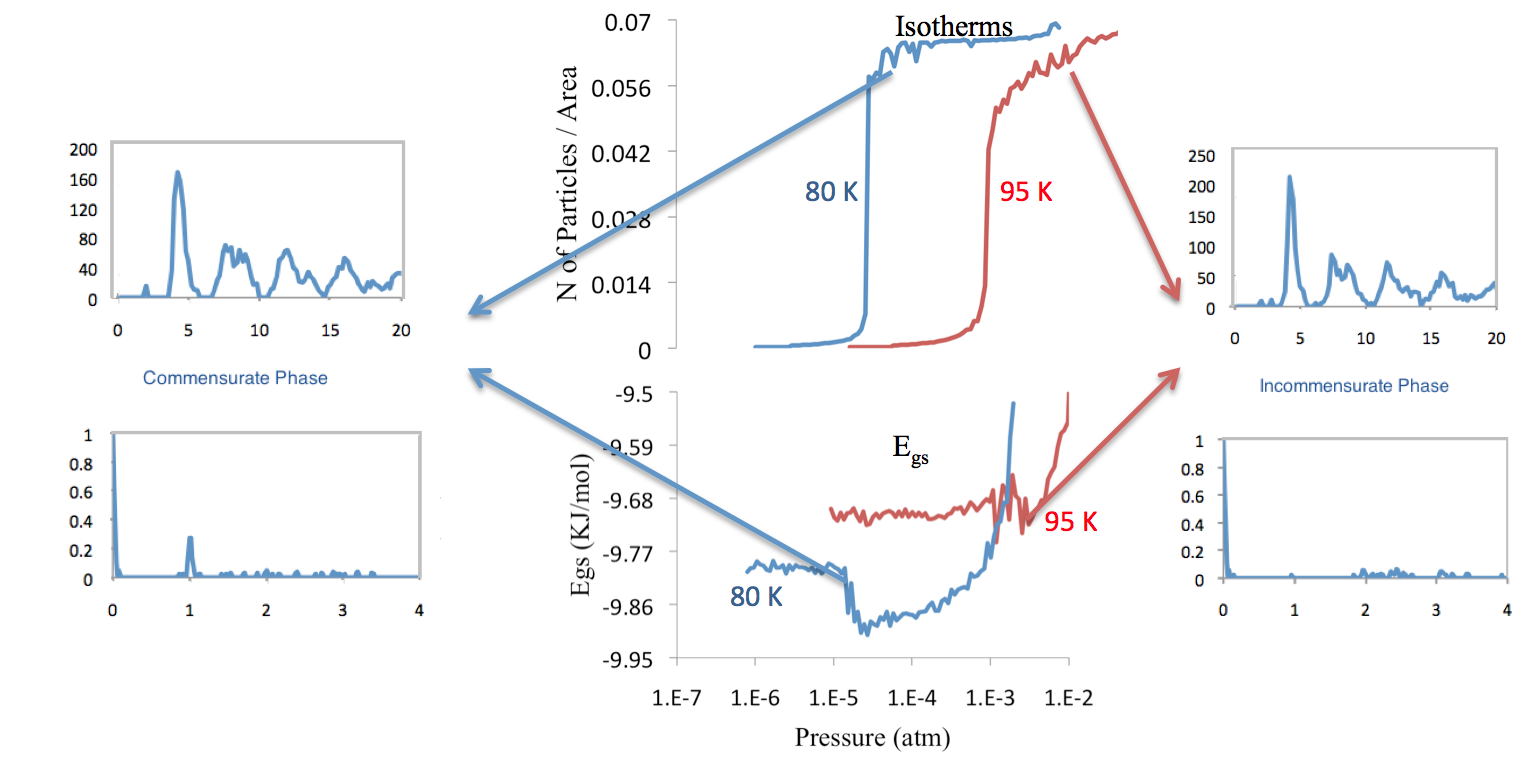}
\caption{Center column: Adsorption isotherms (upper) and gas-surface energy (lower) for Kr at 80 K and 95 K. Left and right columns:  2D-RDR (upper) and SF (lower)  at 80 K (left) and 95 K (right).}
  \label{fig:RDR-SF}
   \end{figure}

 To illustrate our analysis, in \Figref{RDR-SF} we show the adsorption isotherms, gas-surface energy, 2D-RDF and SF  at 80 K and 95 K.  Significant fluctuations in the monolayer density   suggest a change of phase.  The 2D-RDF  in both cases is consistent with a solid phase. 
The SF  $S(\vec{k} = m_1\vec{b_1} + m_2\vec{b_2}$) is shown as a function of $m_1$ (with $m_2=0$).    At $T = 95$ K the SF is mostly flat, meaning that  the phase  is  IS.  At $T = 80$ K, on the other hand, the SF shows  a pick at   $m_1=1$, indicating a structure  consistent with a CS phase. 

\begin{figure}[p]
 \includegraphics[width=5in]{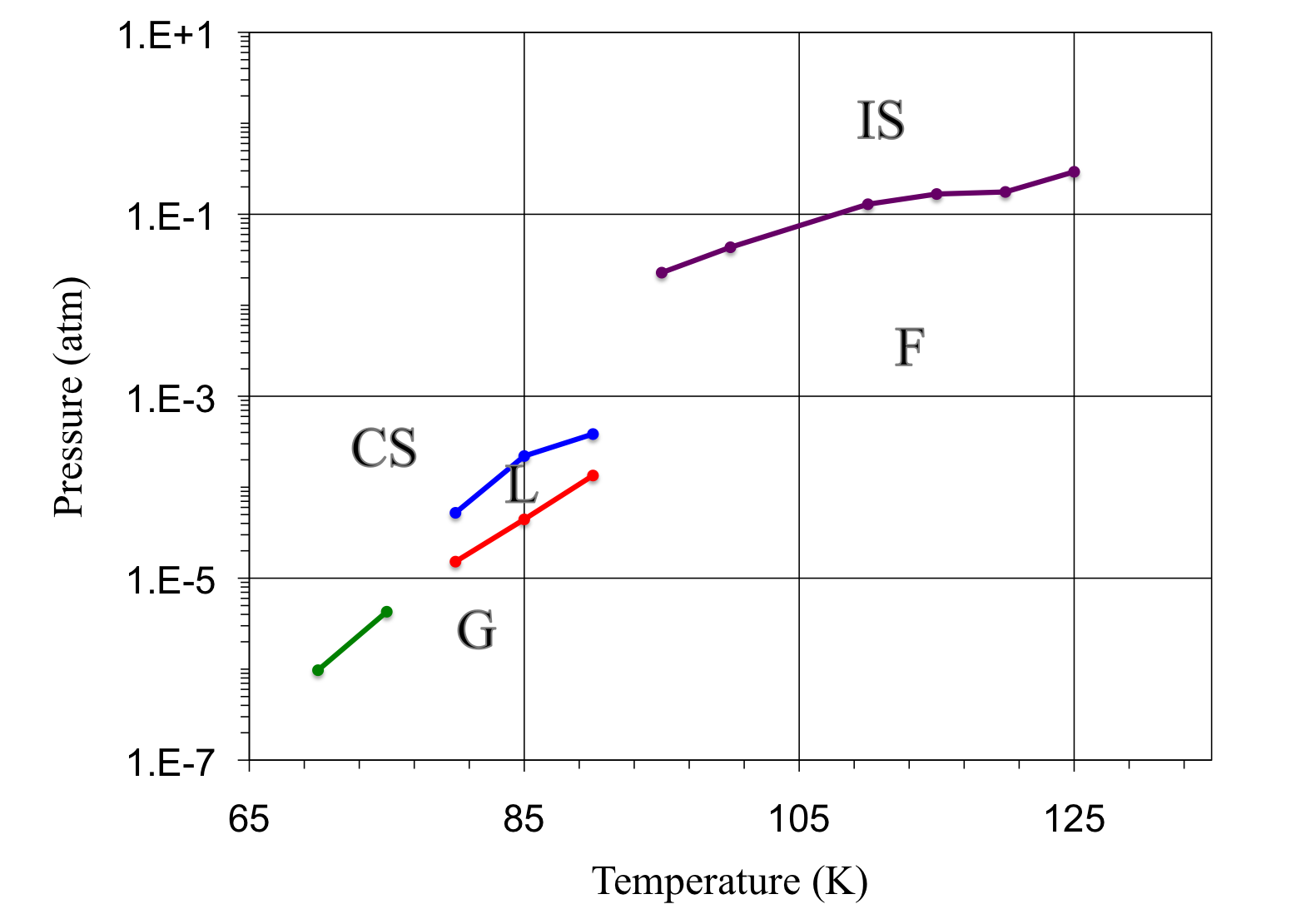}
  \caption{Phase diagram of Kr on monolayer graphene. Labels indicate  incommensurate solid (IS),  commensurate solid (CS), liquid (L), fluid (F) and   gas (G) phases. } 
  \label{fig:phase-Kr}
   \end{figure}

From these analysis   we identified  transitions 2D-G  to 2D-L, 2D-G to 2D-CS and 2D-L to 2D-CS, represented in \Figref{phase-Kr}.  According to our results there is a triple point at temperature  $T_t = 75$ K $\pm 5$ K.

\subsection{\label{sec:level3}  Xenon }

\begin{figure}[p]
 \includegraphics[width=5in]{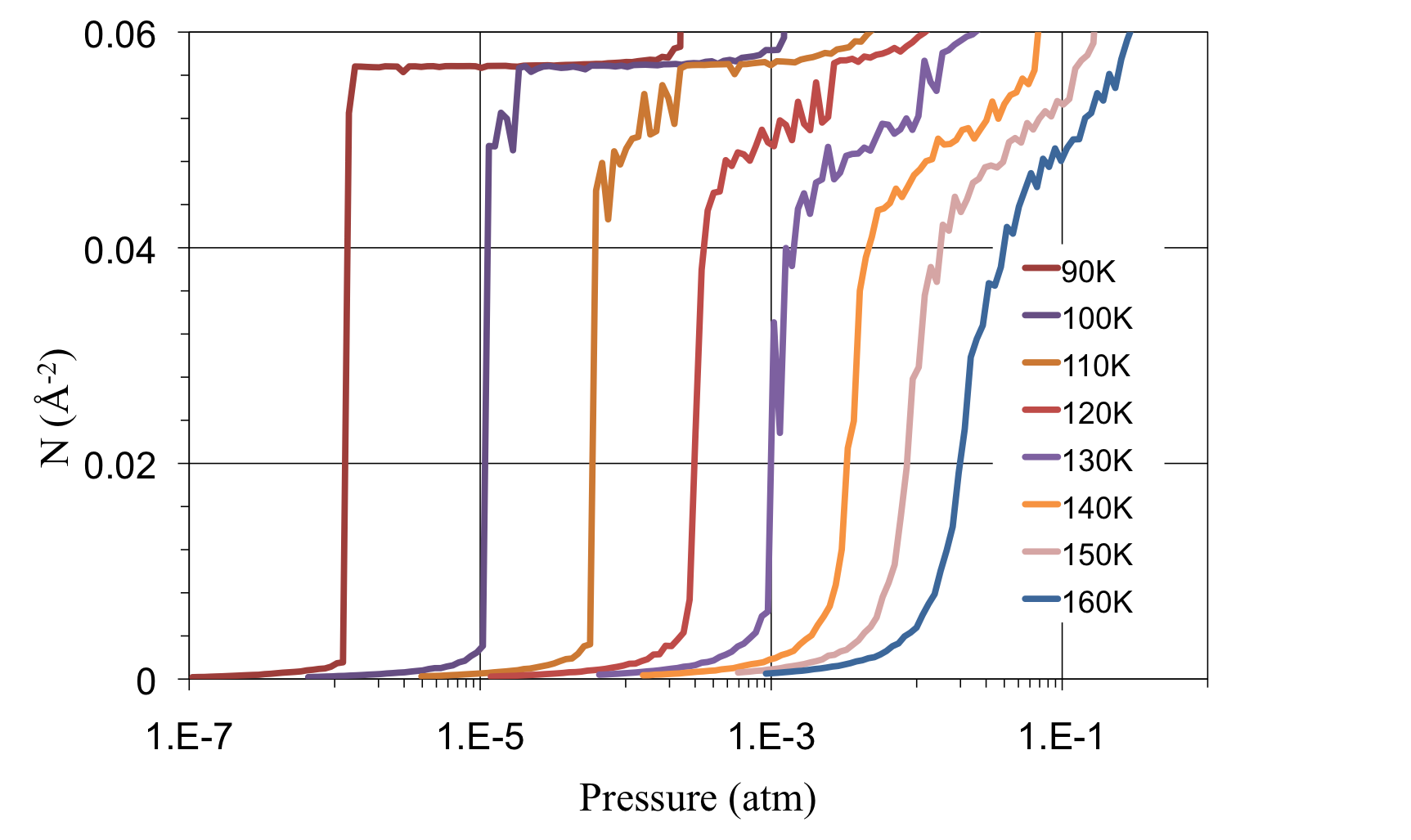}
  \caption{Adsorption Isotherms of Xenon on graphene at temperatures 90K, 100K, 110K, 120K, 130K, and 140K from left to right. }
  \label{fig:Xe}
   \end{figure}

Finally, we simulated the monolayer adsorption of xenon on graphene. We run simulations from T=90K to T=160K. A few adsorptions isotherms  are represented in \Figref{Xe}. 
Our calculation of the 2D G-L critical temperature is $T_c = 115 $ K $\pm 5 $ K, which coincide with the theoretical value, $T_{c(LJ)} =  115$ K and  the experimental value for Xe on graphite, 117 K.  \cite{thomy1981}.

The monolayer density in the simulation is 0.057 \AA$^{-2}$, which is lower than the value of the commensurate solid phase (0.063\AA$^{-2}$) and similar to the LJ solid  lattice (0.055 \AA$^{-2}$).   From the inspection of the 2D-RDF and SF   we observed transitions form  gas and liquid phases to an incommensurate solid phase. 

\begin{figure}[p]
  \includegraphics[width=5in]{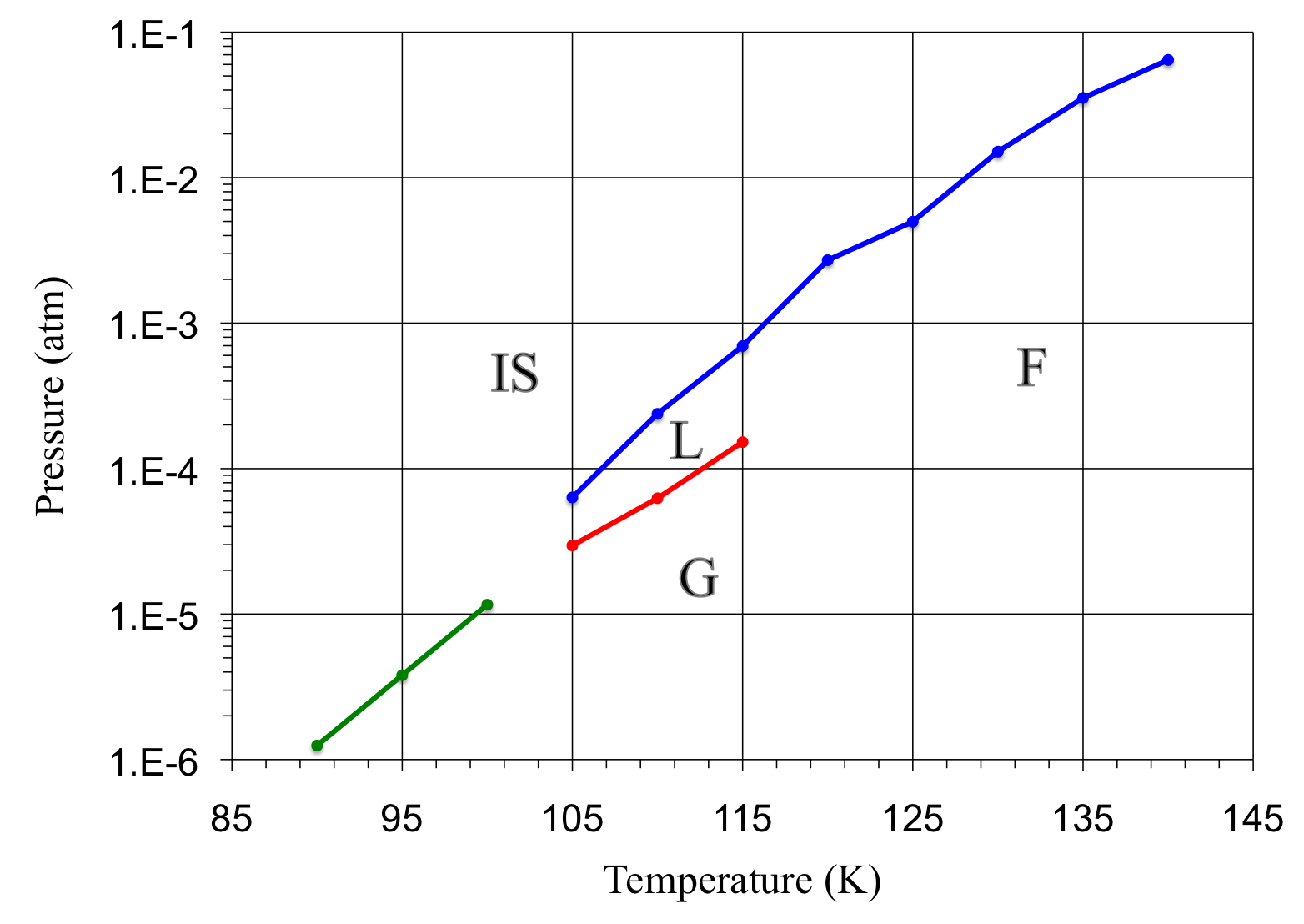}
  \caption{Phase diagram of Xenon on monolayer graphene. Labels correspond to  incommensurate solid (IS), liquid (L)  gas (G) and fluid (F) phases. } 
  \label{fig:phase-Xe}
   \end{figure}

We exhibit the resulting  phase diagram in \Figref{phase-Xe}. We find a triple point at $T_t= 100$ k $\pm 5$ K,  in agreement with the experimental value for Xe on graphite (99 K  \cite{thomy1981}).

     \section{\label{summary}Summary and Conclusions}

Using Grand Canonical Monte Carlo simulations, we studied the adsorption of noble gases (Argon, Krypton, Xenon) on a graphene sheet. 

The main  approximations used in the simulations are that the graphene layer is rigid, and that the effect of the substrate supporting  the graphene is ignored.

Using the adsorption isotherms, we calculated the 2D gas-liquid critical temperature for each gas, resulting in fair agreement with    theoretical  predictions and experimental values of gases on graphite.

We determined the different phases of the gases on graphene by inspecting the radial distribution function and the structure factors. 

We found  2D incommensurate solid phases for krypton, argon and xenon on graphene, and a 2D commensurate solid phase for  krypton   at low temperature. 

\begin{figure}[p]
 \includegraphics[width=\columnwidth]{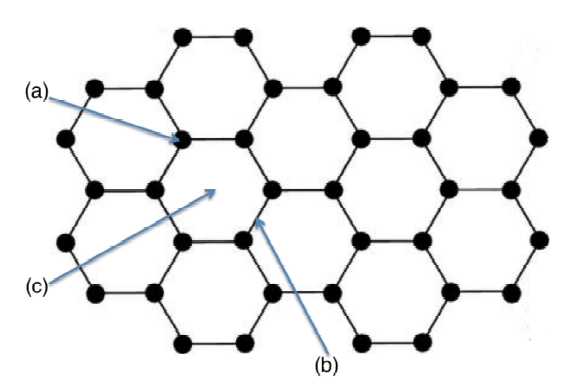}
  \caption{Three different adsorption sites on top of graphene: above (a), bridge (b) and center (c)  }
  \label{fig:site}
   \end{figure}

\begin{figure}[p]
  \includegraphics[width=5in]{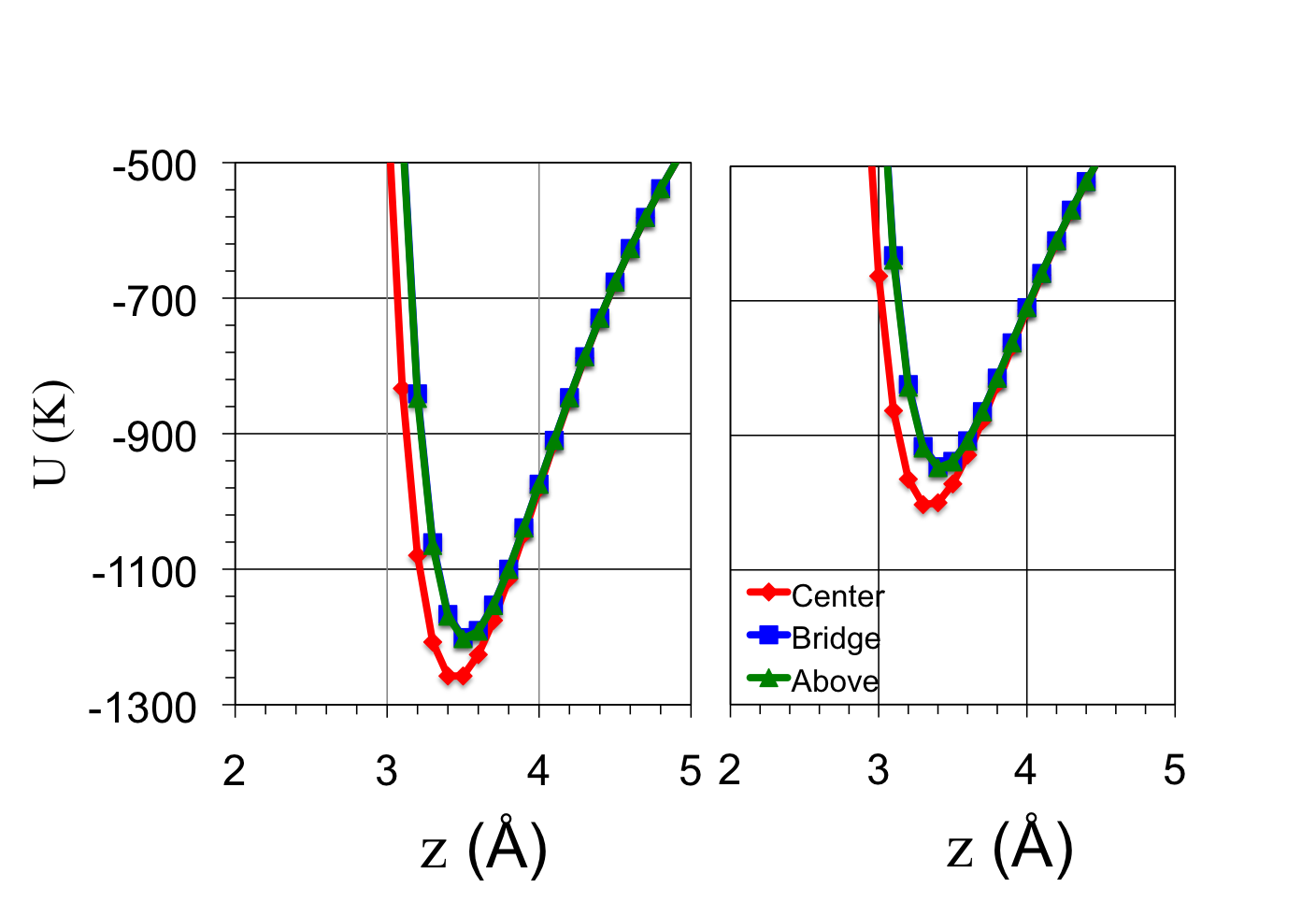}
  \caption{Adsorption potentials of Kr (left) and  Ar (right)  on top of graphene} 
  \label{fig:Pot1}
   \end{figure}

The absence  of a commensurate solid phase in the case of Xe is simply caused by its size: Xe does not "fit" in the 1/6 lattice.  On the other hand,  the qualitative difference between the behavior of Kr and Ar is due to a combination of size and energy effects.  We may have  expected  a higher corrugation of the substrate potential for Kr than for Ar. The corrugation of the potential is defined as the  energy difference between the least attractive and  the most attractive sites on the surface.  Those sites are shown in  \Figref{site}. The most attractive site is on top of the center of a hexagon, and the least attractive sites are both on top of a carbon atom or on top of a bridge.  The corrugation  is 56.7 K  for Kr and 54.9 K for  Ar  (see \Figref{Pot1}). 
These values differ in only  2 K (3\%)! The size of the atoms is also quite similar: Kr is only 6 \% larger than Ar.    However,  Kr seems to have the right size to form a commensurate solid on graphene while almost matching the density of the 2D LJ incommensurate solid,  hence doubling the energy rewards. 

 In conclusion, in spite of the approximations done in our simulations, the comparison  with other available studies is fairly good. The theoretical predictions of the 2D-LJ critical temperature coincide with our values.  The experimental measurements of the critical temperatures agree with our values for Xe,  and are 6 \% and  11  \% lower for Kr and Ar respectively.  This trend suggests that the behavior of the 2D  monolayer is affected by the presence of the substrate,  which is relatively stronger for Ar than for Kr, and less important for Xe. 

We look forward to see reports of  new measurements   for noble gases on graphene and to be able to compare with our results.

\section*{Acknowledgment}
We thank the financial support provided by the National Science Foundation, Center for Integrated Quantum Materials (CIQM), Grant No. DMR-1231319.  We thank the financial support provided by the National Science Foundation, Partnership for Reduced Dimension Materials (PRDM), NSF Grant No. DMR1205608. We thank Milton Cole for his  valuable comments on the manuscript.


\bibliography{mybib.bib}
\end{document}